\documentclass{article}
\title{Point Enclosure Problem for Homothetic Polygons}

\author{Waseem Akram\thanks{E-mail:akram@iitk.ac.in} {\normalsize{ and
 }} Sanjeev Saxena\thanks{E-mail: ssax@iitk.ac.in}\\
Dept. of Computer Science and Engineering,\\ 
Indian Institute of Technology,\\ Kanpur, INDIA-208 016}

\date{\today}
\begin{document}
\maketitle

\begin{center}
    \textbf{Abstract}
\end{center}

In this paper, we investigate the homothetic point enclosure problem:
given a set $S$ of $n$ triangles with sides parallel to three fixed
directions, find a data structure for $S$ that can report all the
triangles of $S$ that contain a query point efficiently. The problem
is ``inverse'' 
of the homothetic range search problem\cite{R2}. We present
an $O(n\log n)$ space solution that supports the queries in $O(\log n
+ k)$ time, where $k$ is the output size. The preprocessing time is
$O(n\log n)$. The same results also hold for homothetic polygons.  

\section{Introduction}

Point enclosure is one of the fundamental problems that has been well
studied in Computational Geometry \cite{R0, R3,
R7, R9}. The problems of this type, typically, are
formulated as follows:
\begin{quote}
Preprocess a set $S$ of input geometrical objects so that given a
query point $q$, the objects of $S$ containing the query point $q$ can
be reported or counted efficiently.
\end{quote} 
In the counting version of the problem, the goal is to report the
number of objects in $S$ containing a query point instead of the
objects themselves. The interval overlapping problem \cite{R0,
R1} and inverse range reporting problem \cite{R1,
R9} are extensively studied problems of this type.

In this paper, we consider the following point enclosure problem.
Given a set $S$ of $n$ triangles of the same shape in the plane,
preprocess $S$ so that a given query point $q$ in the plane, all the
triangles $T$ of $S$ with $T \cap q \ne \phi$ can be reported
efficiently. The trivial way of doing this is to sequentially process
each triangle of $S$ and check whether it contains the query point or
not, which will take linear time. Note that this solution takes
$\Omega(n)$ time in the worst case even when no triangle contains the
query point. We achieve a solution with $O(\log n + k)$ query time,
with $O(n\log n)$ preprocessing time and $O(n\log n)$ space.

Chazelle et al\cite{R2} considered its dual or inverse, namely,
homothetic range search problem. The point enclosure problems with
orthogonal input objects have been well studied\cite{R0, R1, R9}. The
point enclosure problem with input triangles has been studied with
different constraints--- Katz\cite{R6} and Gupta\cite{R4} provided
solutions for the problem when input triangles are fat triangles.
Sharir\cite{R8} proposed a randomized solution for a more general
problem. We are not aware of any work which deals with homothetic
input triangles and provides a deterministic solution. Guting\cite{R5}
gave an optimal time solution for its counting version.

\section{A solution for Right-angled Isosceles Triangles}

First, we consider a simpler variant of the problem in which all the
input triangles are right-angled isosceles triangles with equal sides
parallel to the axes. We present a $O(n\log n)$ space solution for the
variant that supports query in $O(\log n + k)$ time, where $k$ is the
output size. We can obtain a solution for the original homothetic
point enclosure problem for triangles, by using suitable linear
transformation(s).

To obtain a solution for the simpler variant, consider a point $p$ and
a right-angled isosceles triangle $T$ with equal sides parallel to the
exes of the plane. The point $p$ can not be contained by the triangle
$T$ if their horizontal or vertical projections on the axes are
disjoint. We also observe that a triangle whose horizontal projection
contains the $x$-coordinate of point $p$ will contain the point $p$ if
and only if the point $p$ lies between its horizontal side and
hypotenuse. We use these observations to design our solution.

We call the orthogonal projection of a triangle $T$ on the $X$-axis
its $x$-interval, denoted by $T_{h}$. The $y$-interval of a triangle
$T$ is analogously defined which is denoted by $T_{v}$. We denote the
output size by $k(Q)$, where $Q$ is a query point. We omit $Q$ where
the context is clear.

A right-angled triangle is called an axis-parallel right-angled
triangle if its sides incident at the right angle are parallel to
axes.

A triangle $T$ is said to be horizontally closer to a point $p$, if
$T_{h}\cap p_{x}\ne \phi$. The vertical closeness of a triangle for a
point is analogously defined.

Let $S$ be a set of $n$ homothetic right-angled isosceles triangles.
Without loss of generality, we assume that the triangles in the set
$S$ are axis-parallel triangles and present in the first quadrant of
the plane. A point $q = (q_{x}, q_{y})$ lies in a triangle $T$ of $S$
if and only if $q$ is horizontally closer to $T$ and $q_{y}$ lies on
the line segment of the line $x=q_{x}$ inside the triangle $T$.

Based on the $x$-intervals of the triangles of $S$, we construct a
one-dimensional segment tree\cite{R0}, denoted by $\mathcal{T}$.
Each node $v$ of the segment tree $\mathcal{T}$ corresponds to a
closed vertical slab $H(v)$. We say that a triangle $T$ of $S$ crosses
$H(v)$ if $T$ intersects $H(v)$ and none its vertices lie in the
interior of $H(v)$. Let $S(v)$ be the subset of triangles of $S$ that
cross $H(v)$ but do not cross $H(u)$, where $u$ is the parent of $v$
in $\mathcal{T}$. There are $O(\log n)$ nodes $v$ such that $T\in
S(v)$, for a triangle $T$ in $S$. Moreover, any triangle is included
in $O(\log n)$ parts, one in each slab. The size of the segment tree
$\mathcal{T}$ is $O(n\log n)$.

Consider a triangle $T \in S(v)$, for some node $v$. The portion of
$T$ (trapezoid) lying inside the slab $H(v)$ can be partitioned into a
right-angled triangle and a (possibly empty) rectangle. Formally, for
each node $v$ and each triangle $T$ of $S(v)$, we define the
\textbf{trimmed triangle} for $(T, v)$ as the right-angle triangle in
$H(v) \cap T$ that has the points of intersections of the hypotenuse
of $T$ with the vertical sides of the slab $H(v)$ as the endpoints of
its hypotenuse. The remaining portion of $H(v) \cap T$ is a (possibly
empty) rectangle lying below the trimmed triangle. We call it the
\textbf{trimmed rectangle for $(T,v)$}.

Observe that

~\\

{\textbf{Lemma:}}
Let $T$ be a triangle in $S(v)$, for some node $v$ in the segment tree
$\mathcal{T}$. The trimmed triangle for $(T, v)$ is always a
right-angle isosceles triangle and all the trimmed triangles
corresponding to a node $v$ are congruent.

~\\

For each node $v$ in the segment tree $\mathcal{T}$, we associate two
structures. One would return those trimmed triangles of $v$ that
contain the query point while the other one would return the trimmed
rectangles containing the query point. The data structure is built as
follows.

\begin{enumerate}

\item Build a segment tree, denoted by $\mathcal{T}$, for the
$x$-intervals of the triangles present in $S$.

\item If triangle $T\in S(v)$, then compute the trimmed triangle and
rectangle for $(T,v)$. For each trimmed triangle and rectangle, we
maintain a pointer to the defining triangle.

\item For each node $v$ in $\mathcal{T}$:

\begin{enumerate}
\item The trimmed triangles are stored in a list sorted by the
ordinates of their horizontal sides. We denote the sorted list by
$L(v)$.

\item Preprocess the set of the trimmed rectangles corresponding to
$S(v)$ into an interval overlapping structure\cite{R1}, denoted
by $I(v)$.

\end{enumerate}
\end{enumerate}
Queries are processed as follows. Let $q = (q_{x},q_{y})$ be a query
point.
\begin{enumerate}

\item Find the search path for $q_{x}$ in the segment tree
$\mathcal{T}$.

\item For each node $v$ on the search path

\begin{enumerate}

\item[2.1] Find the trimmed rectangles (and hence the corresponding
triangles) of $S(v)$ that contains the point $q$ by querying the
structure $I(v)$ with $q_{y}$ using the method of
Chazelle\cite{R1}.

\item[2.2] Using the binary search in the sorted list, find the
position of $q_{y}$ in the sorted list $L(v)$.

\item[2.3] While moving to the left in the list $L(v)$, report all
the trimmed triangles (and hence corresponding original triangle)
until we get a trimmed triangle for which does not contain the query
point.

\end{enumerate}

\end{enumerate}
The data strucutre can be construcuted in $O(n\log n)$ space and time.
The query time of the algorithm is $O(\log^{2} n + k)$. Using the
fractional cascading technique\cite{R3}, we can improve the
query time by a log factor. Hence, we have the following result.

~\\

{\textbf{Theorem:}}
We can compute a data structure for a given set of $n$ right-angled
isosceles triangles with the same orientation so that, for a given
query point, it can report all the triangles of the input set
containing the query point in $O(\log n + k)$ time. The strucure can
be construcuted in $O(n\log n)$ space and time.

~\\

Using suitable transformations, we can easily obtain a solution for
the problem with homothetic input triangles. The bounds will remain
intact. 

By triangulating input polygons, the result can be extended to the
point enclosure problem for homothetic polygons.

\newpage

\bibliographystyle{acm}

\end{document}